\documentclass[twocolumn,showpacs,preprintnumbers,amsmath,amssymb,subfig]{revtex4}

\usepackage{graphicx}

\begin{document}

\title{Influence of the biquadratic exchange interaction in the classical ground state magnetic response of the antiferromagnetic icosahedron}

\author{N. P. Konstantinidis}
\affiliation{Leoforos Syggrou 360, Kallithea 17674, Athens, Hellas}

\date{\today}

\begin{abstract}
The icosahedron has a ground state magnetization discontinuity in an external magnetic field when classical spins mounted on its vertices are coupled according to the antiferromagnetic Heisenberg model. This is so even if there is no magnetic anisotropy in the Hamiltonian. The discontinuity is a consequence of the frustrated nature of the interactions, which originates in the topology of the cluster. Here it is found that the addition of the next order isotropic spin exchange interaction term in the Hamiltonian, the biquadratic exchange interaction, significantly enriches the classical ground state magnetic response. For relatively weak biquadratic interaction new discontinuities emerge, while for even stronger the number of discontinuities for this small molecule can go up to seven, accompanied by a susceptibility discontinuity. These results demonstrate the possibility of using a small entity like the icosahedron as a magnetic unit whose ground state spin configuration and magnetization can be tuned between many different non-overlapping regimes with the application of an external field.
\end{abstract}


\maketitle

\section{Introduction}
\label{sec:introduction}

The magnetic properties of small molecules have been the subject of intense research activity in the recent decades, as they are attractive candidates for magnetic information storage and as qubits in quantum computers. For example, single molecule magnets exhibit slow relaxation of the magnetization of purely molecular origin and not associated with intermolecular interactions \cite{Gatteschi06,Gatteschi94}. They can be viewed as single bits in a potential quantum computer \cite{Leuenberger01} and demonstrate mesoscopic quantum coherence. More generally, progress in the synthesis of metal complexes has led to the production of nanosized magnetic molecules containing a few dozens of interacting magnetic metal ions, creating the class of molecular nanomagnets \cite{NPK11}. Another way to fabricate small magnetic entities is artificial engineering,
where clusters of magnetic ions are fabricated directly on insulating surfaces and their magnetic properties are measured with scanning tunneling microscopy \cite{Hirjibehedin06,NPK13}. It is thus desirable to classify relatively small molecules according to their magnetic properties, which depend on their topology, the magnitude of the individual spins and the nature of the interactions between them. One can simultaneously search for molecules with the required properties, such as the quantum tunneling of magnetization and slow spin relaxation of single molecule magnets.

Frustrated molecules are of particular interest towards this search. They can be viewed to a good extent as finite analogues of frustrated lattices which have generated extensive research output, especially in the search of the long-sought after spin-liquid phase \cite{Anderson73}. Frustration in finite systems can lead to interesting magnetic phenomena, for example to a discontinuous ground state magnetization in an applied external field, where the molecule's magnetization can be tuned between well-separated values by controlling the field. This is the case when the antiferromagnetic Heisenberg model (AHM) describes the interaction between spins, classical or quantum, mounted on the vertices of fullerene molecules \cite{Coffey92,NPK05,NPK07,NPK16,NPK16-1}. What is surprising in this case is that there are multiple magnetization gaps even if there is no magnetic anisotropy. Their origin lies in the frustrated topology of the spin interactions, following from the special connectivity of the fullerene molecules. They are made up of a number of hexagons proportional to the number of vertices of the molecule, and 12 pentagons. The pentagons introduce frustration and act as defects among the non-frustrated hexagons.
The magnetization response in the presence of frustration has also been considered in similar settings \cite{NPK01,NPK09,Jimenez14,Hucht11,Strecka15,Sahoo12,Sahoo12-1,Nakano14,NPK02,NPK04}.

\begin{figure}
\includegraphics[width=4in,height=3in]{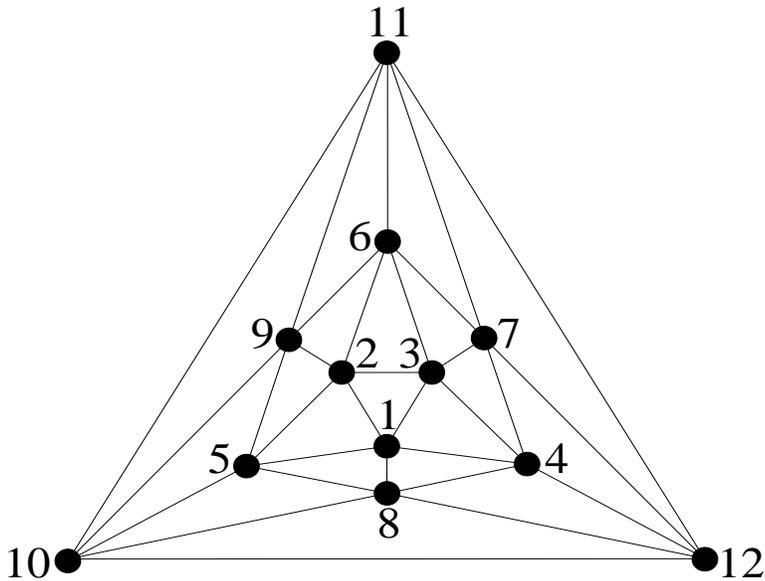}
\caption{Projection of the icosahedron on a plane. The black circles are classical spins $\vec{s}_i$ with magnitude equal to unity, and nearest neighbors interact with bilinear $J$ and biquadratic $J'$ exchange according to the solid lines.
}
\label{fig:icosahedronclusterconnectivity}
\end{figure}

A cluster smaller in size than the smallest fullerene, the dodecahedron, is the icosahedron, with both of them being Platonic solids \cite{Zeyl14}. The icosahedron has 12 equivalent vertices and consists of 20 triangles (Fig. \ref{fig:icosahedronclusterconnectivity}), while it has the spatial symmetry of the icosahedral group $I_h$, the most symmetric group with 120 operations. Due to its triangular structure it has frustrated connectivity, with non-antiparallel neighboring spins in the classical ground state of the AHM \cite{Schmidt03}. Furthermore, for small individual spin quantum number $s$ there are non-magnetic excitations within the singlet-triplet gap and the specific heat has a multi-peak structure as function of temperature, features also of the dodecahedron \cite{NPK05}.

Frustration also results in the icosahedron's classical ground state magnetic response having a discontinuity in an external field within the framework of the AHM \cite{Schroder05}. This has been more specifically attributed to the particular connectivity of the icosahedron, which can be viewed as a strip of a triangular lattice linked with two spins on its top and bottom \cite{NPK15}. These two spins are associated with a magnetization discontinuity when they are free, as they immediately align across an infinitesimal field. This discontinuity survives when the two spins interact with the triangular strip, and it does so until all couplings are equal to each other and the spatial isotropic interaction limit of the icosahedron is recovered. It was also shown that the large degeneracy of the classical ground state leaves its fingerprint down to very small $s$. On the other hand, the classical magnetization discontinuity appears for the first time for a not so low $s=4$.

As in the case of fullerene molecules, the origin of the magnetization discontinuity in the icosahedron lies in its special topology and not in any anisotropy in the spin interactions. With this in mind, in this paper we investigate the classical ground state magnetization response when the next order isotropic exchange term, the biquadratic exchange, is added to the bilinear exchange interaction in the Hamiltonian.
The main goal is to investigate the robustness of the magnetization jump and if more discontinuities appear as a result of the biquadratic interaction. Now there is one more competing term in the Hamiltonian since the bilinear exchange favors antiparallel while the biquadratic exchange perpendicular spins, when the exchange constants for both interactions are positive. Simultaneously the magnetic field tends to align the spins along its axis.
A similar calculation for open odd chains has shown that the biquadratic interaction alters the magnetic response when strong enough \cite{NPK15-1}.

It has been shown that higher order exchange terms are often important for the analysis of experimental data \cite{Furrer11}. A biquadratic exchange interaction term is required for the explanation of the slight deviations from the Land\'e rule for a Mn$^{2+}$ dimer in a CsMgBr$_3$ single crystal \cite{Falk84}. It has also been found important to explain the magnetic susceptibility data of quasiclassical one-dimensional magnetic systems \cite{Gaulin86}. The biquadratic coupling is also important in layered magnetic systems where it can be of considerable strength \cite{Demokritov98}, and has also been proposed to be strong in the vanadium oxide LiVGe$_2$O$_6$ \cite{Millet99}. It has also been realized experimentally by ${}^{23}$Na atoms in optical lattices \cite{Orzel01}, and was invoked to explain the anomalous spin-liquid type properties of NiGa$_2$S$_4$, where it can be quite stronger than the bilinear interaction \cite{Nakatsuji05,Tsunetsugu06,Lauchli06-1}. The biquadratic exchange interaction was derived microscopically by Anderson \cite{Anderson63}. Its importance can not be understated also from the theoretical point of view, for example for the $s=1$ chain \cite{Thorpe72,Haldane83,Haldane83-1,Affleck87,Lauchli06}, where it has also been considered along with biqubic terms for $s=\frac{3}{2}$ \cite{Fridman11}. The influence of the biquadratic interaction has also been investigated in higher dimensions \cite{Hayden10,Kawamura07,Wenzel13}. It has also been invoked to understand the magnetism of the normal state of iron-based supeconductors \cite{Yu15,Zhuo16}. The origin of the biquadratic interaction in the pnictides has been attributed to quantum and thermal fluctuations \cite{Qin14,Fernandes12}.

The main result of the paper is that the addition of the biquadratic coupling not only does not work against the magnetization discontinuity, but more importantly results in a much richer ground state magnetization diagram with respect to magnetization and susceptibility jumps. This is so even though there is no magnetic anisotropy and the icosahedron is only made up of twelve spins. For weak biquadratic interaction a new magnetization jump appears, while increasing the coupling even further brings about further jumps and also susceptibility discontinuities. As stated in the previous paragraph the biquadratic can be even quite stronger than the bilinear coupling. When the two are comparable in magnitude the ground state magnetization curve of the icosahedron has a maximum number of seven magnetization and one susceptibility discontinuities, a formidable number for such a small molecule that far supersedes the single discontinuity of the AHM. Discontinuities survive even when the biquadratic is much stronger than the bilinear interaction. These results demonstrate the potential of using the icosahedron as a small entity with tunable ground state magnetization between many different spin configurations via the application of an external field.


The plan of this paper is as follows: In Sec. \ref{sec:model} the Hamiltonian of the bilinear-biquadratic exchange model is introduced, and in Sec. \ref{sec:zeromagneticfield} its lowest energy configuration is calculated for zero magnetic field. Section \ref{sec:nonzeromagneticfield} extends the investigation to non-zero magnetic fields, while Sec. \ref{sec:conclusions} presents the conclusions.

\section{Model}
\label{sec:model}

The icosahedron has $N=12$ equivalent vertices that are five-fold coordinated, and consists of 20 triangles (Fig. \ref{fig:icosahedronclusterconnectivity}). It has the spatial symmetry of the icosahedral group $I_h$. We consider classical spins $\vec{s}_i$ which are unit vectors located on its vertices, and they interact according to the Hamiltonian of the bilinear-biquadratic exchange model:


\begin{equation}
H = \sum_{<ij>} [ J \vec{s}_i \cdot \vec{s}_j + J' ( \vec{s}_i \cdot \vec{s}_j )^2 ] -  h \sum_{i=1}^{N} s_i^z
\label{eqn:model}
\end{equation}


$<ij>$ indicates that there are interactions only between nearest neighbors $i$ and $j$. The first term is the bilinear exchange interaction, which has strength $J$. The next term is the biquadratic exchange, with strength $J'$. A magnetic field $\vec{h}$ is also added and is taken to point without loss of generality along the $z$ direction. The bilinear exchange favors antiparallel nearest-neighbor spins for positive $J$, while the biquadratic perpendicular when $J'$ is positive. These two relative orientations can not be satisfied for every bond even when each coupling is considered alone in zero magnetic field, the reason being the frustrated topology of the cluster. The situation is further complicated by the Zeeman term of Hamiltonian (\ref{eqn:model}), through which the spins gain maximum magnetic energy by aligning themselves with the field. The magnetic properties are determined by the competition between these three terms and by the frustrated connectivity of the cluster. The exchange interactions are parametrized as $J$=cos$\omega$ and $J'$=sin$\omega$. Here we are interested in the lowest energy configuration of Hamiltonian (\ref{eqn:model}) for non-negative $J$ and $J'$, or $0 \leq \omega \leq \frac{\pi}{2}$. The calculations were done numerically \cite{Coffey92,NPK05,NPK07,NPK15,NPK15-1,NPK16,NPK16-1}. Each spin $\vec{s}_i$ is a classical unit vector defined by a polar $\theta_i$ and an azimuthal $\phi_i$ angle. A random initial configuration is chosen and each angle is moved opposite the direction of its gradient, until the global minimum of the energy is reached. The procedure is repeated for different initial configurations and the same lowest energy configuration is produced each time.

The lack of quantum fluctuations leads to additional degeneracies in the classical configurations (see for example the inset of Fig. 5 in Ref. \cite{NPK15}). Here the symmetry of the lowest classical energy configuration is determined from the requirement that a spatial symmetry operation does not change the correlation between any two nearest-neighbor spins. The symmetry groups that leave the correlations unchanged are subgroups of $I_h$ \cite{Altmann94}. The classical lowest energy configurations of the icosahedron are degenerate, and whenever they are presented it is for one of the degenerate configurations, or colorings, while all the colorings can be decomposed according to $I_h$ \cite{Axenovich01,NPK15}. The saturation magnetic field is given by $h_{sat}=(5+\sqrt{5})(J+2J')$ \cite{NPK15-1,comment}.

\section{Zero Magnetic Field}
\label{sec:zeromagneticfield}

When $h=0$ the lowest energy configuration of Hamiltonian (\ref{eqn:model}) is determined solely from the competition of the bilinear and biquadratic exchange terms. For $\omega=0$ only the bilinear exchange is non-zero and the Hamiltonian is reduced to the one of the AHM. In its lowest energy configuration nearest-neighbor spins are not antiparallel due to frustration, and the nearest-neighbor correlation equals $-\frac{\sqrt{5}}{5}$ for every pair \cite{Schmidt03}. The symmetry group of the lowest energy configuration (Fig. \ref{fig:icosahedronCFsixdifferent}(a)) is $I_h$. Spins come in groups of three with the same polar angle, with these angles adding up to $\pi$ in two pairs \cite{NPK15}.
The azimuthal angles within each of the four groups differ by $\frac{2\pi}{3}$, and all azimuthal angles can be chosen to be multiples of $\frac{\pi}{3}$.

\begin{figure}
\centerline{
\includegraphics[width=2in,height=1.5in]{correnergyicosahedronbiquadratic}
\includegraphics[width=1.8in,height=1.5in]{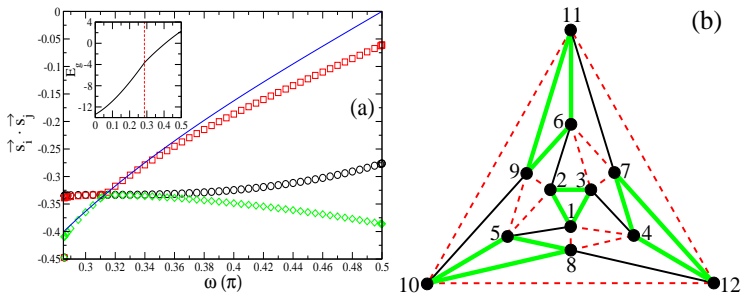}
}
\caption{(Color online) (a) The (black) circles, (red) squares, and (green) diamonds show the unique nearest-neighbor correlations $\vec{s}_i \cdot \vec{s}_j$ in the lowest energy configuration of Hamiltonian (\ref{eqn:model}) for zero field as a function of $\omega$=arctan$\frac{J'}{J}$. For $\omega \leq 0.28565 \pi$ all correlations are equal to $-\frac{1}{\sqrt{5}}$. When $\omega=\frac{\pi}{2}$ the nearest-neighbor correlations can be either ferromagnetic or antiferromagnetic, but only the negative values are shown. The (blue) line plots the nearest-neighbor correlation $-\frac{1}{2 \textrm{tan}\omega}$ of an open odd chain for comparison. The (black) line in the inset shows the lowest energy $E_g$ for the whole $\omega$ range. The (red) dashed line is defined by $\omega=0.28565 \pi$, and for smaller $\omega$ the energy is given by $-6 ( \sqrt{5} J - J' )$. (b) The unique nearest-neighbor correlations for $\omega > 0.28565 \pi$ shown on the icosahedron: the (black) lines correspond to the (black) circles of (a), the (red) dashed lines to the (red) squares, and the (green) thick lines to the (green) diamonds.
}
\label{fig:icosahedronzerofield}
\end{figure}

When $\omega \neq 0$ Hamiltonian (\ref{eqn:model}) can be rewritten as a sum of perfect squares plus a constant term \cite{NPK15-1}, however unlike the case of an open chain it is not possible to directly deduce the lowest energy configuration, since the squares can not be minimized individually due to the frustrated topology of the icosahedron. Numerical minimization shows that the ground state configuration remains the one of the AHM for $\omega \leq 0.28565 \pi$.
For bigger $\omega$ the ground state nearest-neighbor correlation is not any more the same for every pair, but assumes three different values (Fig. \ref{fig:icosahedronzerofield}). The symmetry group of the lowest energy configuration is the cubic group $T$. The line in Fig. \ref{fig:icosahedronzerofield}(a) shows the nearest-neighbor correlation $-\frac{1}{2 \textrm{tan}\omega}$ of an open odd chain for comparison \cite{NPK15-1}. For $\omega$=arctan$\frac{3}{2}$ or $\frac{J'}{J}=\frac{3}{2}$ all nearest-neighbor correlations are equal to $-\frac{1}{3}$ coinciding with the value of the open odd chain, nullifying the effect of frustration. For $\omega$$<$arctan$\frac{3}{2}$ the open odd chain has stronger antiferromagnetic nearest-neighbor correlations, while the opposite is true for bigger $\omega$.

When $\omega=\frac{\pi}{2}$ only the biquadratic exchange is non-zero in Hamiltonian (\ref{eqn:model}), and each ground state nearest-neighbor correlation can be antiferromagnetic or ferromagnetic with the same magnitude. This results in a degenerate ground state manifold which corresponds to a range of total magnetization $M$ values, just like the case of the open odd chain for $\omega >$arctan$\frac{1}{2}$ \cite{NPK15-1}. These values range from zero to $\frac{M}{N}=0.56012$ for the total magnetization per spin. For $\omega=\frac{\pi}{2}$ there is also a special relationship between the nearest-neighbor correlation values. The sum of the squares of the ones with the minimum and maximum magnitude equals twice the square of the third correlation function.

The inset of Fig. \ref{fig:icosahedronzerofield}(a) shows the ground state energy as a function of $\omega$, which for $\omega \leq 0.28565 \pi$ equals $-6 ( \sqrt{5} J - J' )$. An estimate of the importance of frustration is given by comparing the value of $\omega$ where the AHM lowest energy configuration ceases to be the ground state, with the corresponding value of an open odd chain. In the latter case the value is quite smaller and equals arctan$\frac{1}{2}=0.14758 \pi$ \cite{NPK15-1}, which shows that in the case of the icosahedron frustration makes the AHM ground state configuration more robust against the biquadratic exchange interaction.


\section{Non-Zero Magnetic Field}
\label{sec:nonzeromagneticfield}



\begin{figure}
\includegraphics[width=3.6in,height=2.7in]{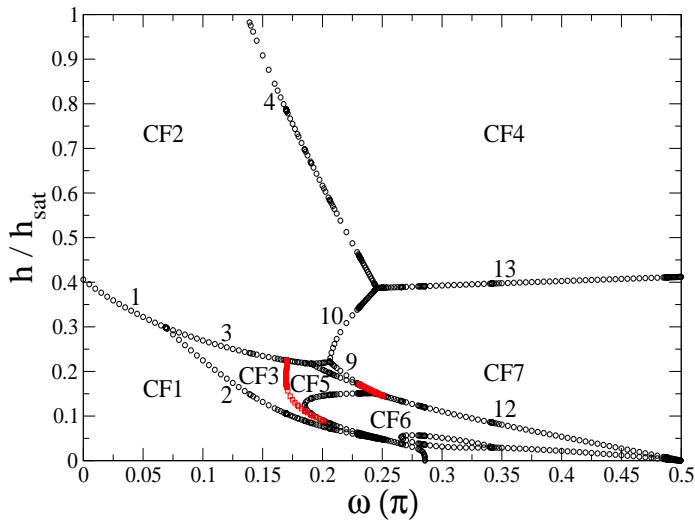}
\caption{(Color online) The (black) circles show the location of the magnetization and the (red) squares of the susceptibility discontinuities as a function of $\omega$=arctan$\frac{J'}{J}$ and $\frac{h}{h_{sat}}$. The magnetization discontinuities are distinguished by different numbers and the susceptibility discontinuities by different primed numbers according to Table \ref{table:magnsuscdisc}. CFi indicates the lowest energy configuration for the different areas of the plot, with i=$1,\dots,9$.
}
\label{fig:icosahedronconfigurations}
\end{figure}
\begin{figure}
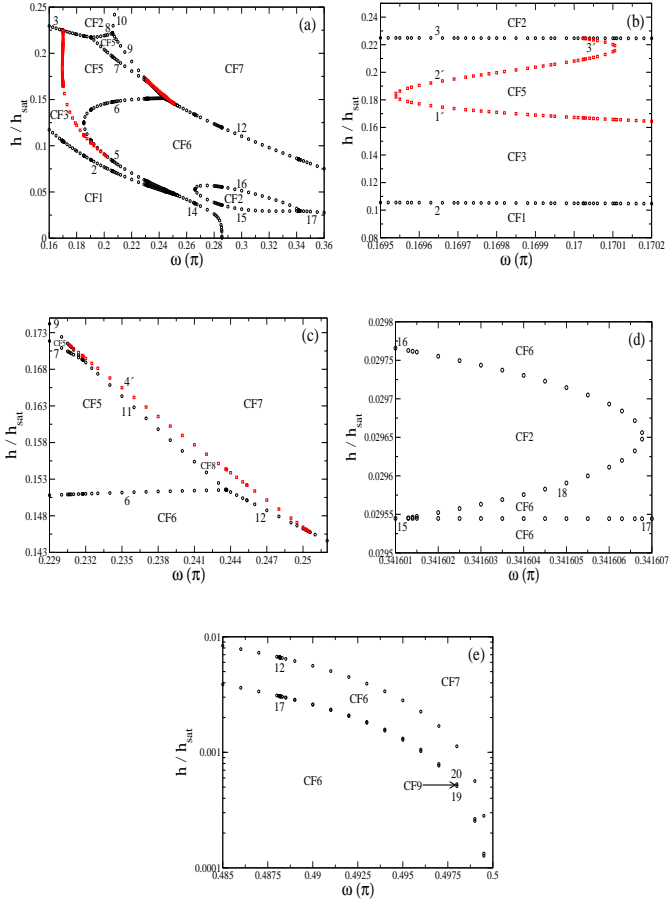

\centerline{
\includegraphics[width=0.24\textwidth,height=1.4in]{icosahedronconfig1.eps}
\includegraphics[width=0.24\textwidth,height=1.4in]{icosahedronconfig2.eps}
}
\vspace{17pt}
\centerline{
\includegraphics[width=0.24\textwidth,height=1.4in]{icosahedronconfig3.eps}
\hspace{0pt}
\includegraphics[width=0.24\textwidth,height=1.4in]{icosahedronconfig4.eps}
}
\vspace{17pt}
\centerline{
\includegraphics[width=0.24\textwidth,height=1.4in]{icosahedronconfig5.eps}
}
\caption{(Color online) Specific parts of Fig. \ref{fig:icosahedronconfigurations} in greater detail.
}
\label{fig:icosahedronconfigurationsfivedifferent}
\end{figure}
\begin{figure}
\includegraphics[width=2.6in,height=3.5in,angle=270]{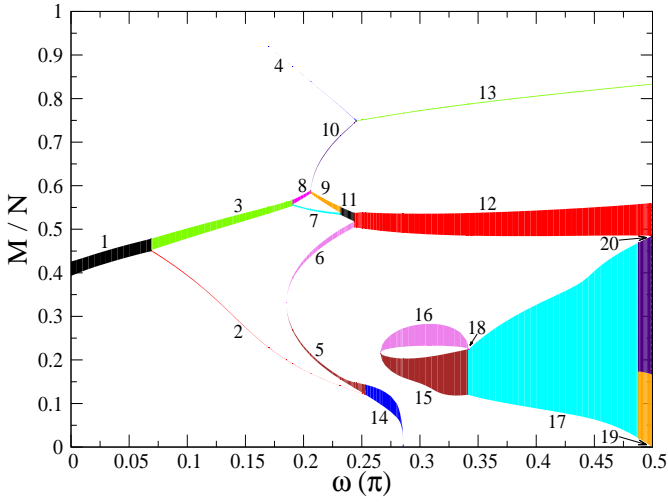}
\caption{(Color online) Inaccessible classical ground state magnetizations per spin $\frac{M}{N}$ as a function of $\omega$=arctan$\frac{J'}{J}$. They originate in the different magnetization discontinuities and are distinguished by a different number according to Table \ref{table:magnsuscdisc} (each of the ones numbered 1 to 10 have a different color, with the same color pattern followed by the ones numbered 11 to 20).
}
\label{fig:icosahedrondiscontinuities}
\end{figure}
\begin{figure}
\includegraphics[width=4in,height=3in]{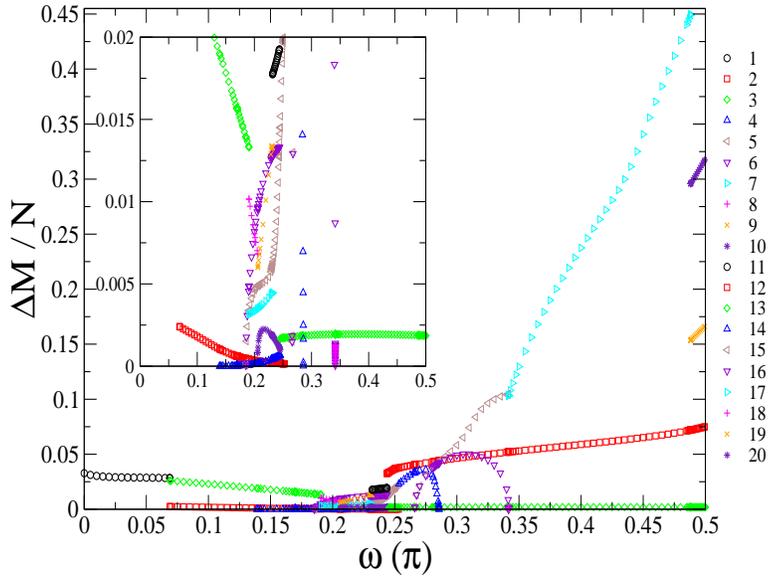}
\caption{(Color online) Magnetization change per spin $\frac{\Delta M}{N}$ for the classical ground state magnetization discontinuities as a function of $\omega$=arctan$\frac{J'}{J}$. Each discontinuity is distinguished by a different number (and color), in accordance with Fig. \ref{fig:icosahedrondiscontinuities}.
}
\label{fig:icosahedrondiscontinuitieswidth}
\end{figure}

A non-zero magnetic field in Hamiltonian (\ref{eqn:model}) results in a multitude of magnetization and susceptibility discontinuities. Their locations as functions of $\omega$ and $\frac{h}{h_{sat}}$ are plotted in Fig. \ref{fig:icosahedronconfigurations} and in finer detail in Fig. \ref{fig:icosahedronconfigurationsfivedifferent}. Each magnetization discontinuity is distinguished by an index, while each susceptibility discontinuity by a primed index. The total number of different lowest energy spin configurations CFi is nine, i=$1,\dots,9$. The values of $\omega$ and $\frac{h}{h_{sat}}$ where discontinuities appear or disappear are given in Table \ref{table:magnsuscdisc}, along with the total number of magnetization and susceptibility discontinuities $N_M$ and $N_{\chi}$ right after the listed $\omega$ values. The inaccessible magnetizations are highlighted in Fig. \ref{fig:icosahedrondiscontinuities}, and the corresponding magnetization gap widths are plotted in Fig. \ref{fig:icosahedrondiscontinuitieswidth}. The bulk of the discontinuities occurs for relatively low magnetic fields.

\begin{figure}
\centerline{
\includegraphics[width=0.24\textwidth,height=1.4in]{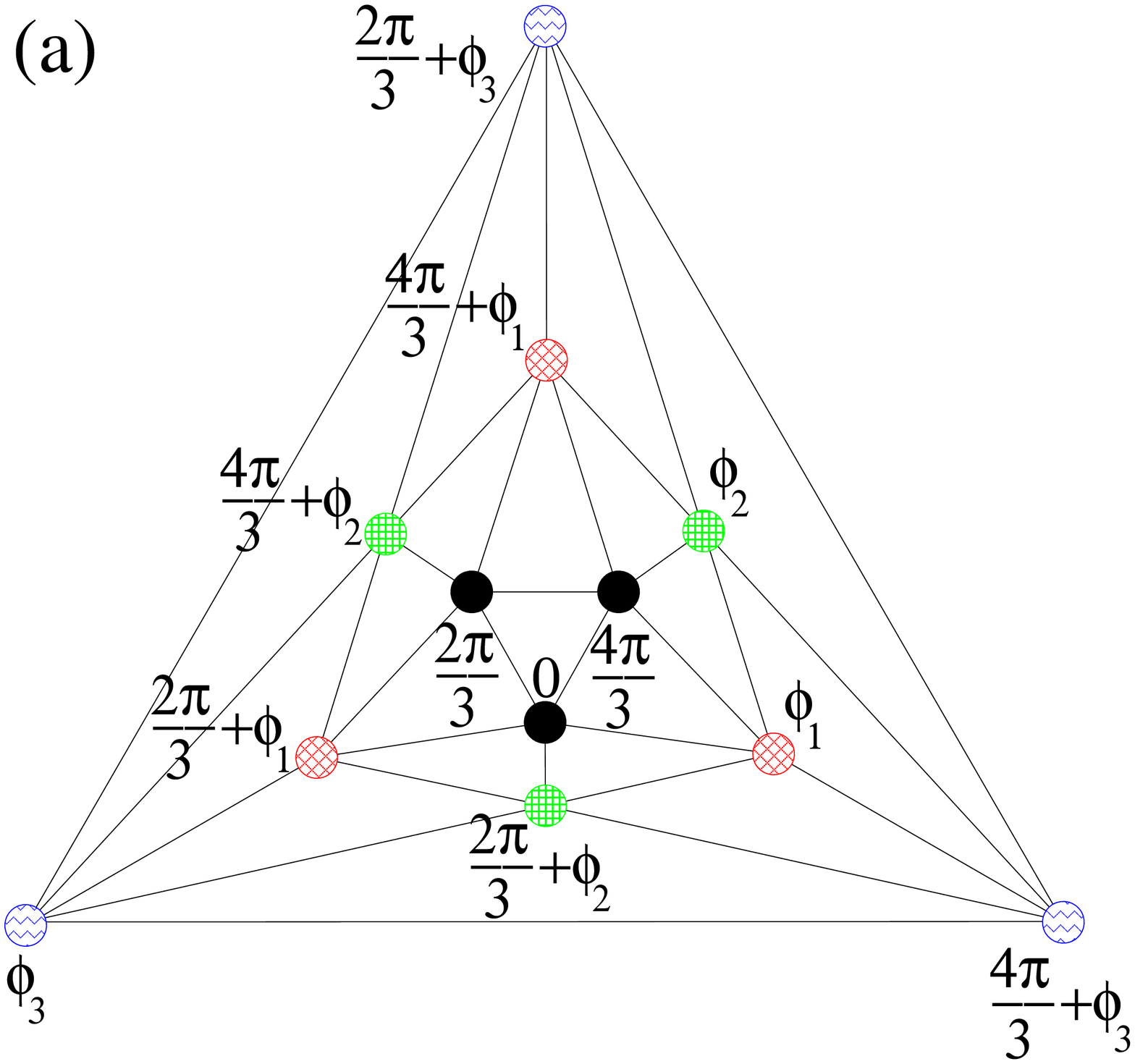}
\includegraphics[width=0.24\textwidth,height=1.4in]{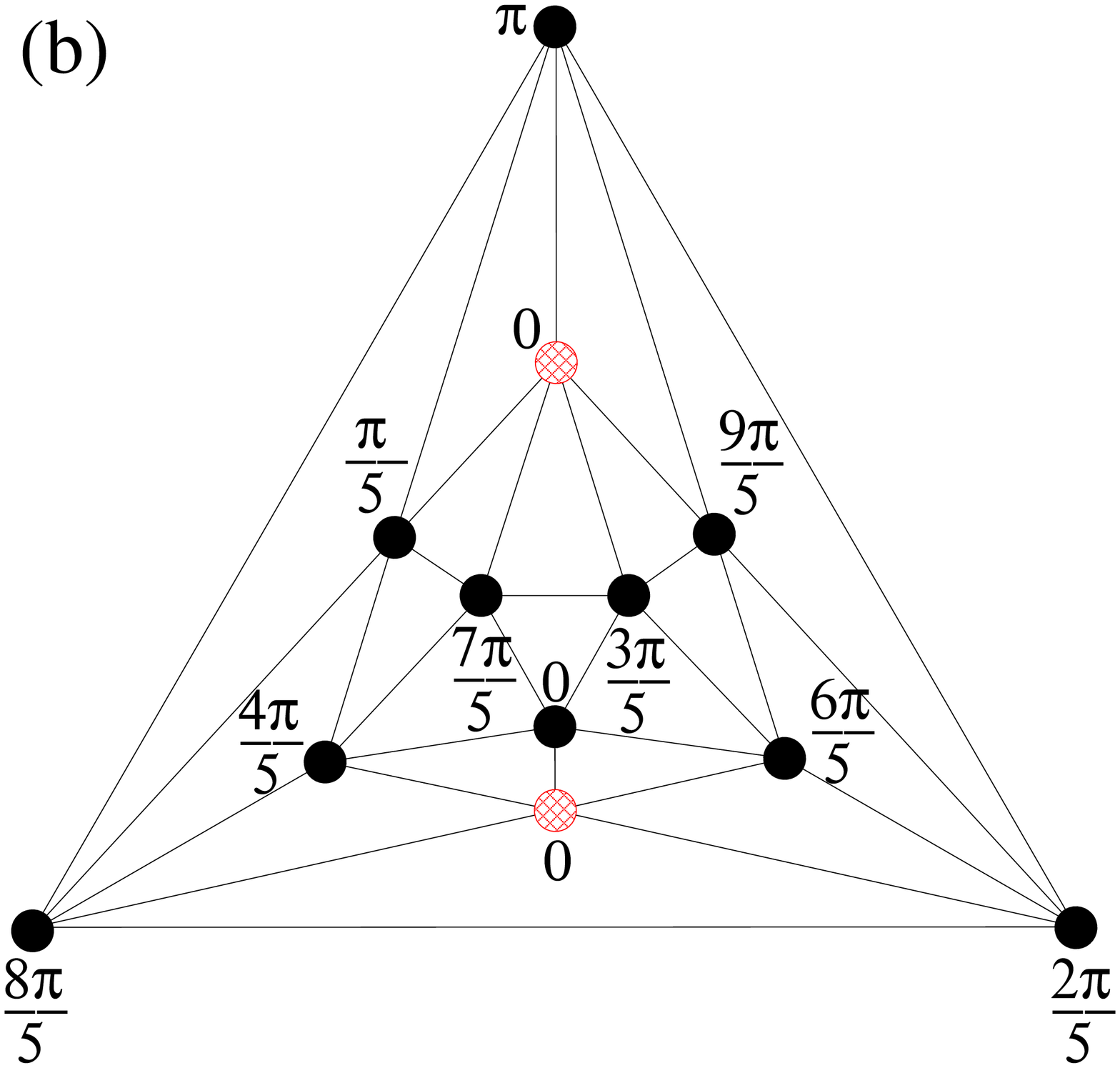}
}
\vspace{17pt}
\centerline{
\includegraphics[width=0.24\textwidth,height=1.4in]{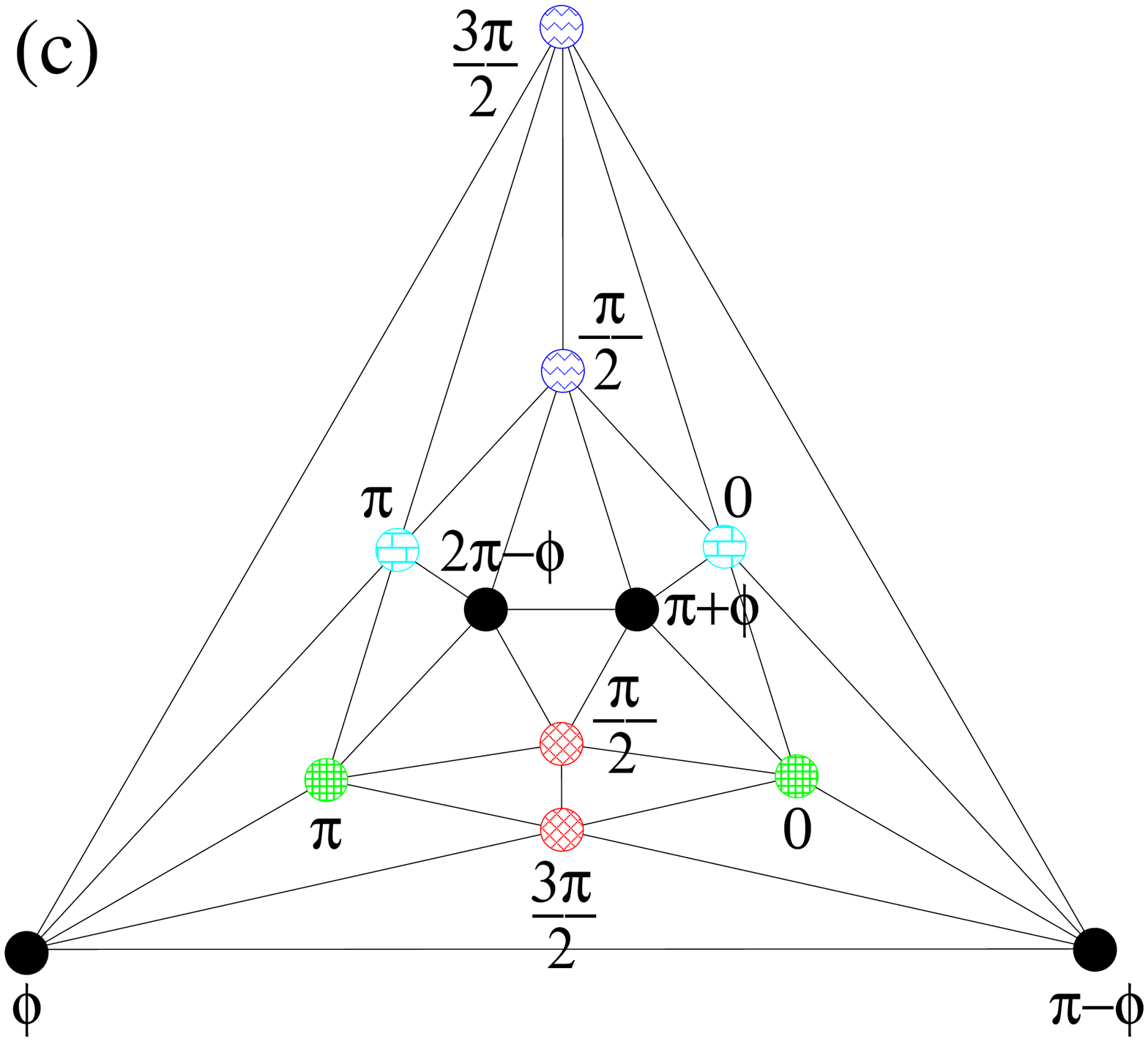}
\hspace{0pt}
\includegraphics[width=0.24\textwidth,height=1.4in]{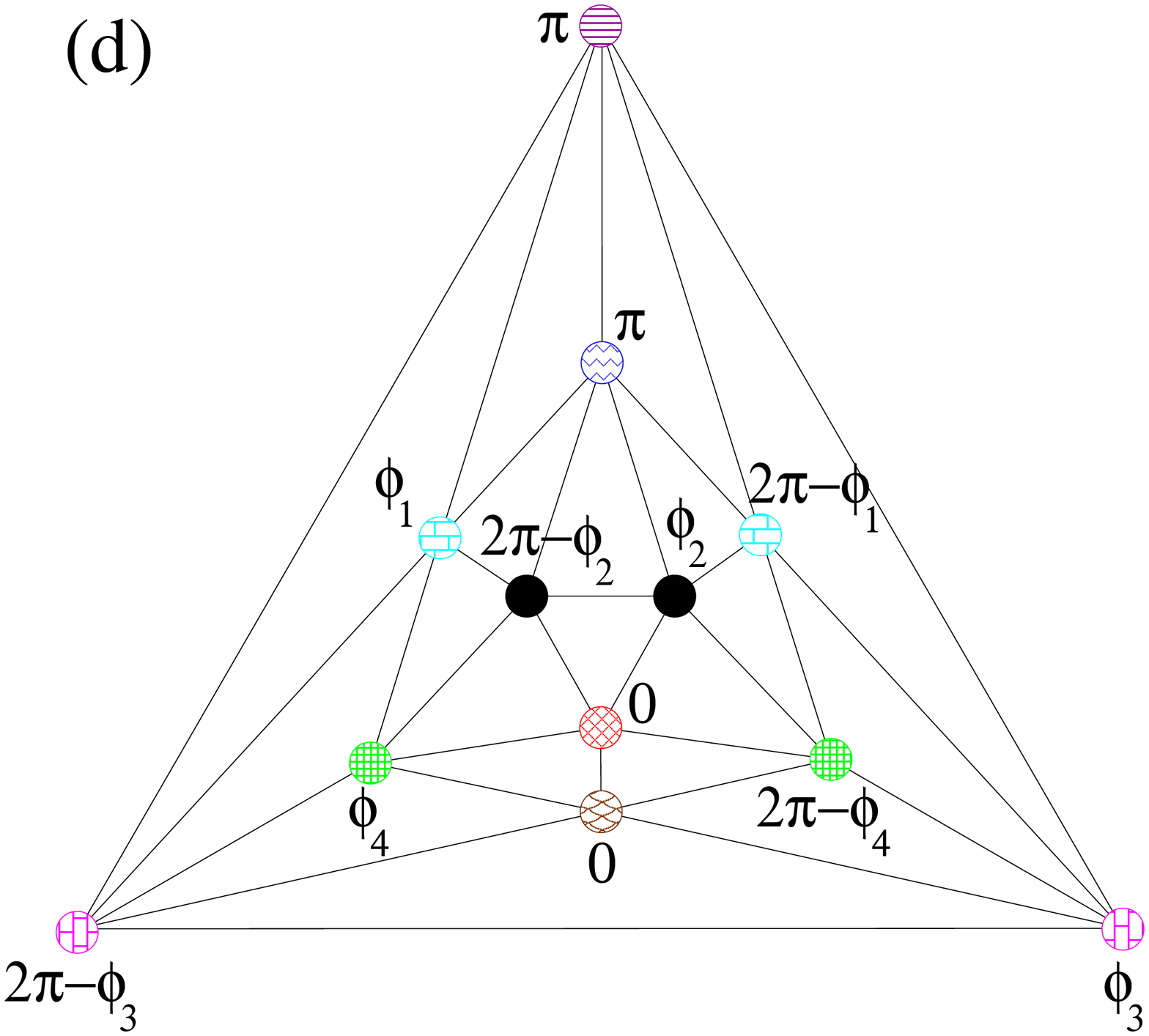}
}
\vspace{17pt}
\centerline{
\includegraphics[width=0.24\textwidth,height=1.4in]{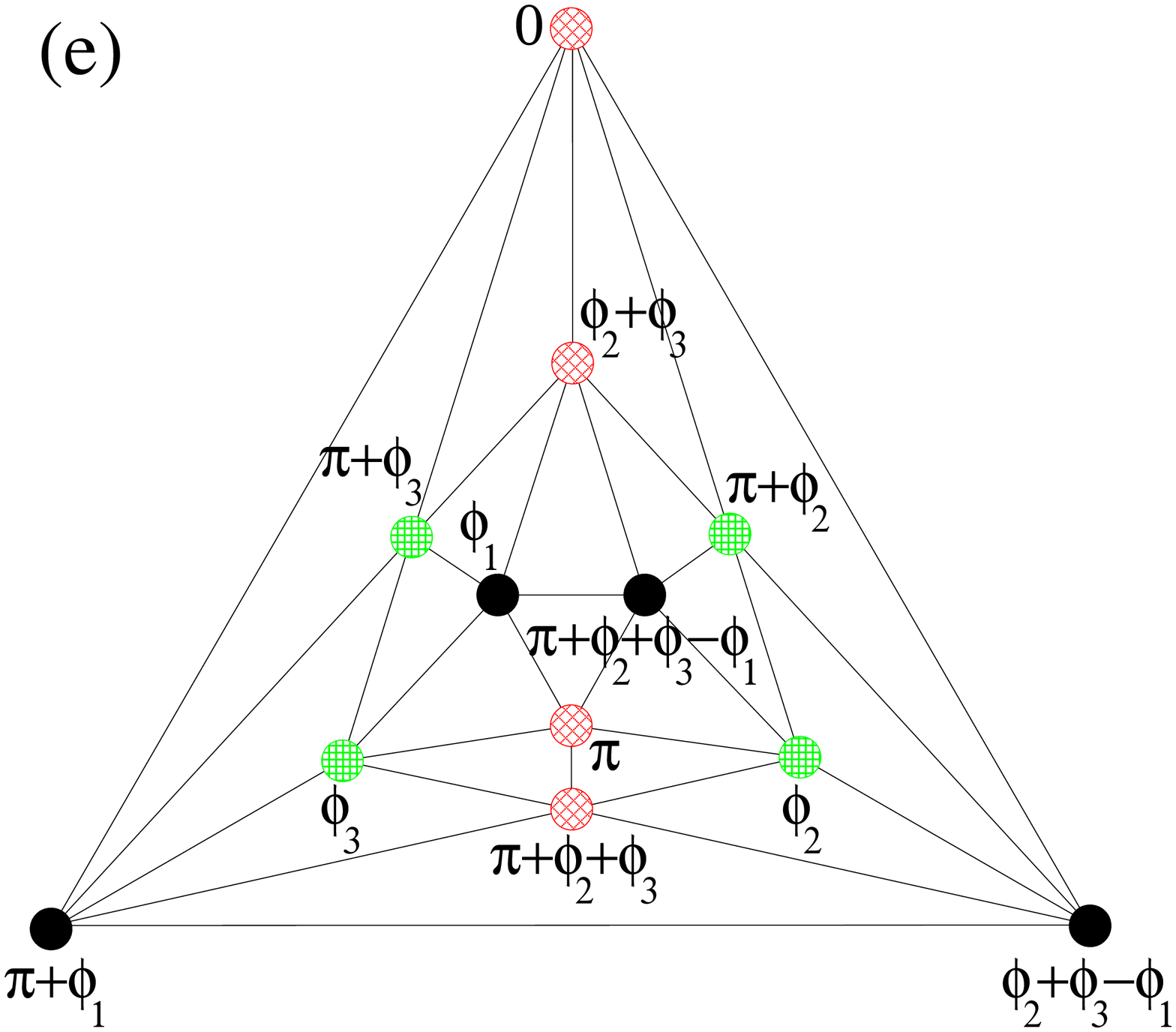}
\hspace{0pt}
\includegraphics[width=0.24\textwidth,height=1.4in]{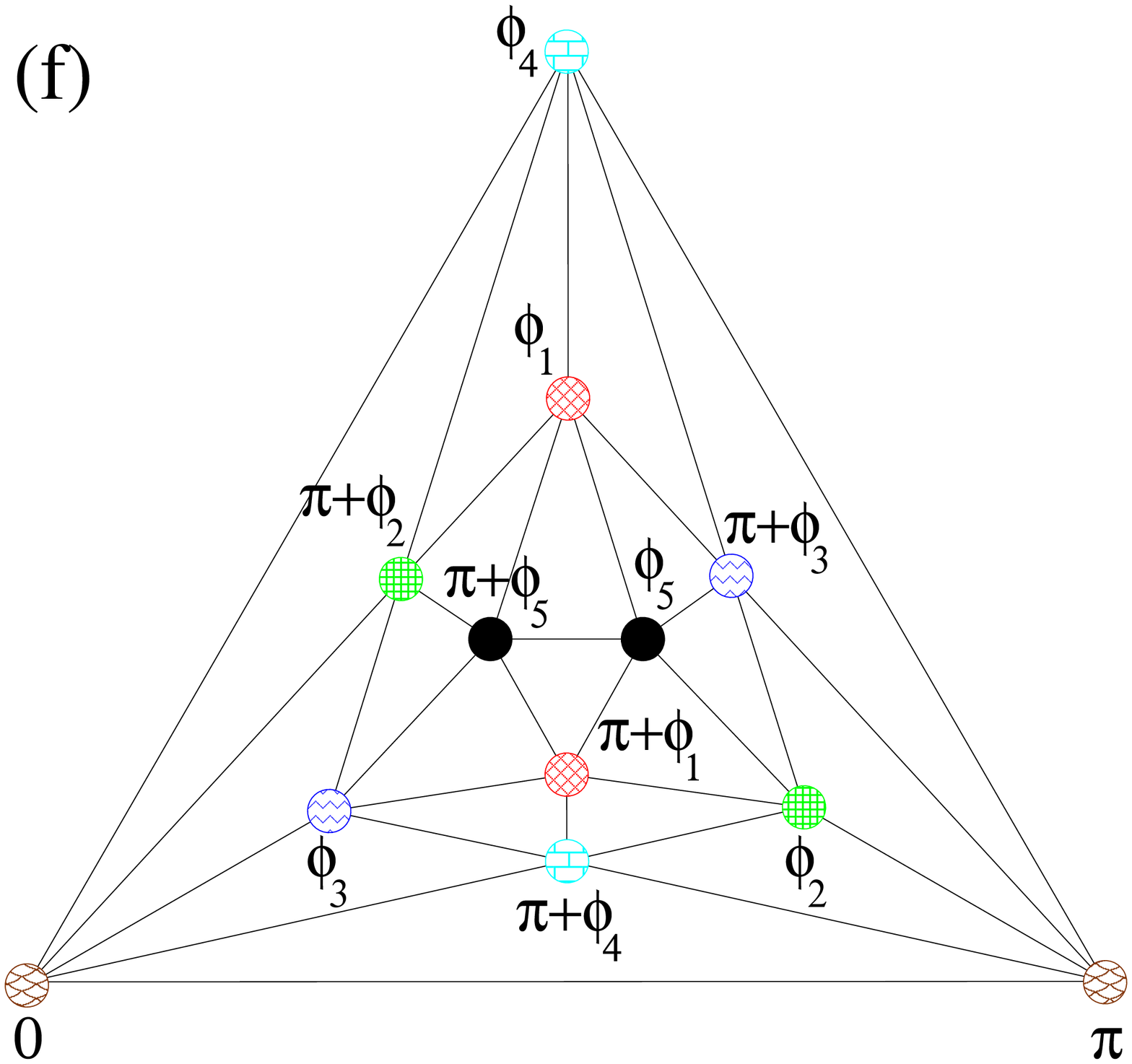}
}
\caption{(Color online) Projection of the icosahedron on a plane (see Fig. \ref{fig:icosahedronclusterconnectivity} for description). Circles of the same pattern (and color) represent equal polar angles in the different lowest energy configurations. The azimuthal angles are explicitly given in terms of $\pi$ and different independent parameters. (a) In configuration CF6 the azimuthal angles are expressed by three independent parameters $\phi_i$, $i=1,2,3$. The symmetry of this configuration is $C_3$. When $\phi_1=\phi_2=\phi_3=0$ the configuration reduces to CF1, whose symmetry is $C_{3v}$. If furthermore the magnetic field is zero the distinct polar angles add up to $\pi$ in two pairs, and the symmetry is $I_h$. Configuration CF4 is derived from configuration CF1 by setting the polar angles equal in pairs, so that spins 4 to 9 have a common polar angle, while the rest of the spins another one. Its symmetry is $D_{3d}$. (b) In configuration CF2 the two (red) spins with the crossed pattern have zero polar angle. The azimuthal angles for the rest of the spins are multiples of $\frac{\pi}{5}$. The symmetry of this configuration is $D_{5d}$. (c) In configuration CF3 the azimuthal angles are expressed by a single parameter $\phi$. Its symmetry is $C_{2v}$. (d) Configuration CF5 has four parameters $\phi_i$, $i=1,\dots,4$, and symmetry $C_s$. (e) Configuration CF7 has three parameters $\phi_i$, $i=1,2,3$, and symmetry $D_2$. (f) Configuration CF8 has five parameters $\phi_i$, $i=1,\dots,5$, and symmetry $C_2$.
}
\label{fig:icosahedronCFsixdifferent}
\end{figure}

In the absence of biquadratic exchange ($\omega=0$) in Hamiltonian (\ref{eqn:model}) $M$ is discontinuous at $\frac{h}{h_{sat}}=0.40603$ \cite{Schroder05,NPK15}. Magnetization jump 1 survives for small $\omega$ (Fig. \ref{fig:icosahedronconfigurations}) and is associated with ground state configurations CF1 and CF2. CF1 (Fig. \ref{fig:icosahedronCFsixdifferent}(a)) is derived from the $\omega=0$ zero field configuration (Sec. \ref{sec:zeromagneticfield}) by precluding any pair of the four unique polar angles to add up to $\pi$, and has $C_{3v}$ symmetry. In CF2 (Fig. \ref{fig:icosahedronCFsixdifferent}(b)) two spins are aligned with the field axis, while the rest share the polar angle and their azimuthal angles assume ten different equidistant values. The symmetry of this configuration is $D_{5d}$.

\begin{figure}
\includegraphics[width=4in,height=3in]{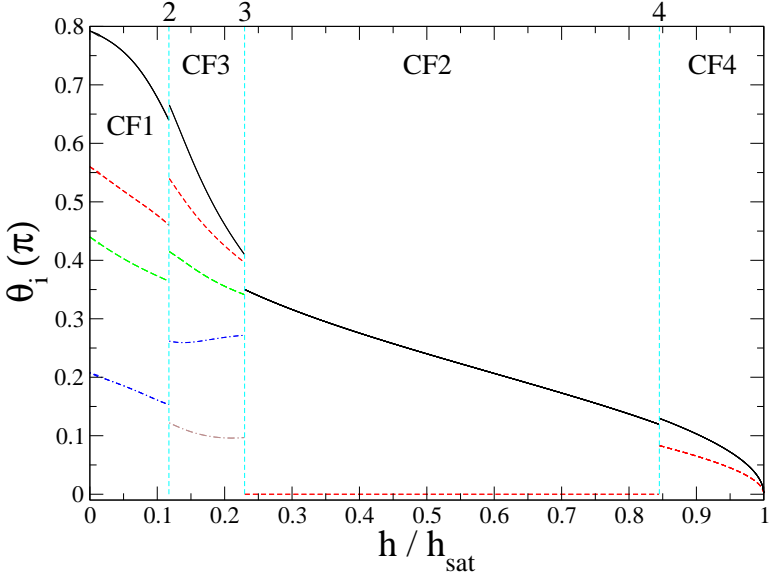}
\caption{(Color online) Polar angles $\theta_i(\pi)$ as a function of $\frac{h}{h_{sat}}$ in the lowest energy configuration of Hamiltonian (\ref{eqn:model}) for $\omega=0.16 \pi$. The dashed (cyan) vertical lines show the location of the magnetization discontinuities 2, 3, and 4, which are at $\frac{h}{h_{sat}}=0.11734$, 0.22955, and 0.84495 (Fig. \ref{fig:icosahedronconfigurations}). The saturation field $h_{sat}=13.31302$.
}
\label{fig:icosahedronomega=0.16PIangles}
\end{figure}

Discontinuity 1 splits in discontinuities 2 and 3 for the relatively small value $\omega=0.068897\pi$ ($\frac{J'}{J}=0.21989$) (Fig. \ref{fig:icosahedronconfigurations}). Inbetween the two jumps the ground state configuration is CF3 (Fig. \ref{fig:icosahedronCFsixdifferent}(c)). The symmetry is reduced with respect to its two neighboring configurations, being $C_{2v}$. Four of the spins have a common polar angle and their azimuthal angles are expressed by a single parameter. The rest of the spins share polar angles in pairs. For each pair the azimuthal angles differ by $\pi$, and all the azimuthal angles of these pairs can be chosen to be multiples of $\frac{\pi}{2}$. For $\omega=0.137 \pi$ ($\frac{J'}{J}=0.459$) (Fig. \ref{fig:icosahedronconfigurations}) discontinuity 4 emerges from saturation, together with lowest energy configuration CF4 for high fields. CF4 is derived from CF1 by setting the polar angles equal in pairs (Fig. \ref{fig:icosahedronCFsixdifferent}(a)), and has $D_{3d}$ symmetry. Fig. \ref{fig:icosahedronomega=0.16PIangles} shows the dependence of the unique polar angles on the ratio of the magnetic field over its saturation value $\frac{h}{h_{sat}}$ for $\omega=0.16 \pi$, where magnetization discontinuities 2, 3 and 4 are present. The polar angles turn gradually towards the field, even though their field dependence is not necessarily monotonic.

When $\omega=0.1695 \pi$ ($\frac{J'}{J}=0.5893$) susceptibility discontinuities 1' and 2' emerge (Fig. \ref{fig:icosahedronconfigurationsfivedifferent}(b)). Inbetween them the lowest energy configuration is CF5 (Fig. \ref{fig:icosahedronCFsixdifferent}(d)). It is up to now the least symmetric classical ground state of the icosahedron, having $C_s$ symmetry. Each of the four spins along the central line has its own polar angle, while their azimuthal angles are either zero or $\pi$. Spins placed symetrically with respect to the central line have the same polar angle, and their azimuthal angles add up to $2\pi$. A third susceptibility discontinuity 3' appears for $\omega=0.1700 \pi$ ($\frac{J'}{J}=0.5914$) (Fig. \ref{fig:icosahedronconfigurationsfivedifferent}(b)).

When $\omega=0.18519 \pi$ ($\frac{J'}{J}=0.65773$) magnetization discontinuities 5 and 6 appear (Fig. \ref{fig:icosahedronconfigurationsfivedifferent}(a)) along with configuration CF6 (Fig. \ref{fig:icosahedronCFsixdifferent}(a)). In it the spins are divided in four groups of three with a common polar angle, and within each group the azimuthal angles differ by $\frac{2\pi}{3}$. Not all azimuthal angles can be chosen as multiples of $\frac{2\pi}{3}$ and three independent parameters determine the configuration in the azimuthal plane. Configuration CF6 has $C_3$ symmetry.

\begin{figure}
\includegraphics[width=4in,height=3in]{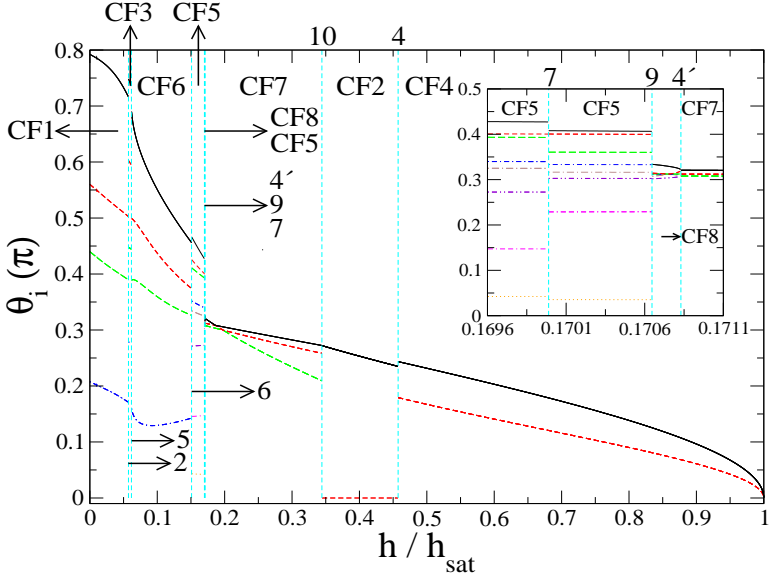}
\caption{(Color online) Polar angles $\theta_i(\pi)$ as a function of $\frac{h}{h_{sat}}$ in the lowest energy configuration of Hamiltonian (\ref{eqn:model}) for $\omega=0.231 \pi$. The dashed (cyan) vertical lines show the location of the magnetization discontinuities 2, 5, 6, 7, 9, 10, and 4, which are at $\frac{h}{h_{sat}}=0.057262$, 0.061697, 0.15094, 0.16999, 0.17065, 0.34428, and 0.45735, and of the susceptibility discontinuity 4', which is at $\frac{h}{h_{sat}}=0.17083$ (Figs. \ref{fig:icosahedronconfigurations}, \ref{fig:icosahedronconfigurationsfivedifferent}(a) and \ref{fig:icosahedronconfigurationsfivedifferent}(c)). The saturation field $h_{sat}=15.01745$. The inset shows the region with the three discontinuities which are close to each other in more detail.
}
\label{fig:icosahedronomega=0.231PIangles}
\end{figure}

\begin{figure}
\includegraphics[width=4in,height=3in]{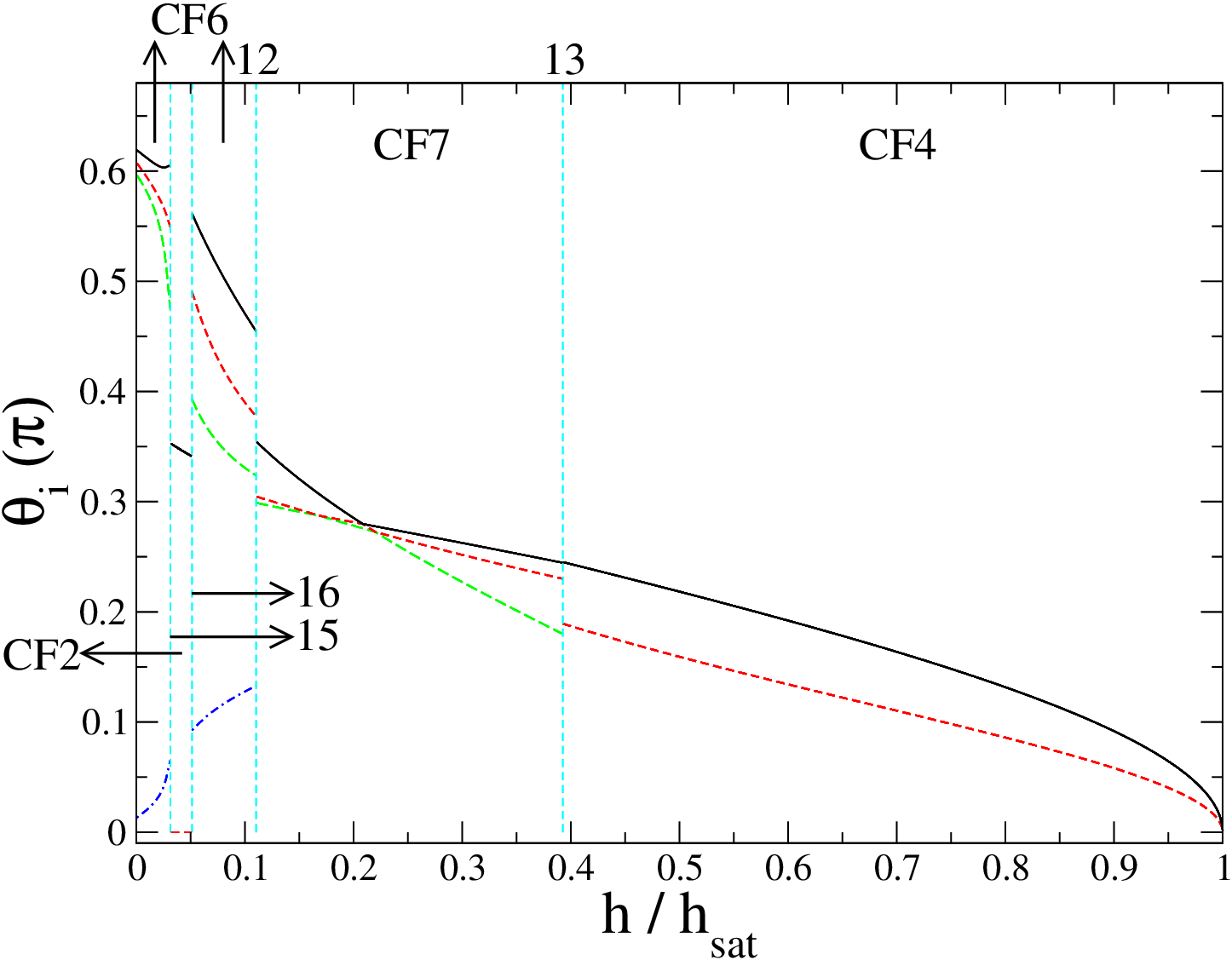}
\caption{(Color online) Polar angles $\theta_i(\pi)$ as a function of $\frac{h}{h_{sat}}$ in the lowest energy configuration of Hamiltonian (\ref{eqn:model}) for $\omega=0.3 \pi$. The dashed (cyan) vertical lines show the location of the magnetization discontinuities 12, 13, 15, and 16, which are at $\frac{h}{h_{sat}}=0.031395$, 0.051341, 0.11033, and 0.39265 (Figs. \ref{fig:icosahedronconfigurations} and  \ref{fig:icosahedronconfigurationsfivedifferent}(a)). The saturation field $h_{sat}=15.96146$.
}
\label{fig:icosahedronomega=0.3PIangles}
\end{figure}

At $\omega=0.20605 \pi$ ($\frac{J'}{J}=0.75599$) discontinuity 8 splits in discontinuities 9 and 10 (Fig. \ref{fig:icosahedronconfigurationsfivedifferent}(a)), and the ground state inbetween them is configuration CF7 (Fig. \ref{fig:icosahedronCFsixdifferent}(e)). It has three groups of four spins having the same polar angle and azimuthal angles differing in pairs by $\pi$, with a total of three independent parameters determining the azimuthal angles. Its symmetry is $D_2$. Configuration CF8 (Fig. \ref{fig:icosahedronCFsixdifferent}(f)) appears for the first time with the susceptibility discontinuity 4' (Fig. \ref{fig:icosahedronconfigurationsfivedifferent}(c)) for $\omega=0.2306 \pi$ ($\frac{J'}{J}=0.8850$). Its symmetry is $C_2$, and has as low symmetry as configuration CF5 does. Spins come in pairs with the same polar angle and azimuthal angles differing by $\pi$. Starting at this value of $\omega$ the number of discontinuities is maximum, with a total of seven magnetization and one susceptibility jumps. This persists up to $\omega=0.23178 \pi$ ($\frac{J'}{J}=0.89161$) where discontinuities 7 and 9 merge into discontinuity 11 (Fig. \ref{fig:icosahedronconfigurationsfivedifferent}(c)). The maximum number of discontinuities thus occurs when the biquadratic is roughly 90\% of the bilinear interaction. Fig. \ref{fig:icosahedronomega=0.231PIangles} shows the dependence of the unique polar angles on $\frac{h}{h_{sat}}$ for $\omega=0.231 \pi$ as the various discontinuities are encountered. Figure \ref{fig:icosahedronomega=0.3PIangles} does the same for $\omega=0.3 \pi$.

When the biquadratic exchange becomes very strong there are four magnetization discontinuities (Figs. \ref{fig:icosahedronconfigurations} and \ref{fig:icosahedronconfigurationsfivedifferent}(e)). In this region the magnetization jumps are the widest (Figs. \ref{fig:icosahedrondiscontinuities} and \ref{fig:icosahedrondiscontinuitieswidth}). Configuration CF9 appears for $\omega=0.48818 \pi$ ($\frac{J'}{J}=26.917$) and possesses no symmetries, with each spin having its own polar and azimuthal angle. For vanishing bilinear exchange the three low-field discontinuities disappear. For zero field nearest-neighbor spins have the possibility to have positive or negative correlation of the same magnitude, which results in a degenerate manifold for the ground state (Sec. \ref{sec:zeromagneticfield}). The non-accessible lower magnetizations when $\omega$ is close to $\frac{\pi}{2}$ become part of the zero field degenerate manifold for $\omega=\frac{\pi}{2}$, which an open odd chain has for tan$\omega > \frac{1}{2}$ \cite{NPK15-1}. This results in only a single magnetization discontinuity for $\omega=\frac{\pi}{2}$.

\section{Conclusions}
\label{sec:conclusions}

The icosahedron consists of only 12 vertices and has the highest possible point group symmetry. It has been shown that its frustrated connectivity results in a classical ground state magnetization discontinuity in an external field when spins mounted on its vertices interact according to the AHM, and that this discontinuity survives down to $s=4$ \cite{Schroder05,NPK15}. Since the discontinuous magnetic response originates in the connectivity of the icosahedron and not in any anisotropy in the interactions, in this paper the ground state magnetization was calculated when the next order isotropic spin exchange interaction term, the biquadratic exchange interaction, is added to the Hamiltonian. It was found that inclusion of this term significantly enriches the classical ground state magnetic response, with such a small entity supporting so many discontinuities. The maximum number of discontinuities is seven of the magnetization and one of the susceptibility when the biquadratic is roughly 90\% of the bilinear exchange coupling. This corresponds to a small magnet that can be tuned between many lowest energy spin configurations with different magnetizations by application of an external field. It is a central goal to characterize the magnetic response and more generally the strongly correlated electronic behavior of molecular nanomagnets according to their topology, the nature of the interactions between their spins, the value of $s$, and also the degree of itinerancy of the spins.

\bibliography{icosahedron}

\begin{thebibliography}{53}
\bibitem[{\citenamefont{\textrm{The saturation field value has also been
  verified by a small angle expansion close to saturation for both of the two
  high-field lowest energy configurations of the icosahedron.}}()}]{comment}
\bibinfo{author}{\bibnamefont{\textrm{The saturation field value has also been
  verified by a small angle expansion close to saturation for both of the two
  high-field lowest energy configurations of the icosahedron.}}}
\end{thebibliography}

\newpage

\begin{table}
\begin{center}
\caption{$\omega$=arctan$\frac{J'}{J}$ values in units of $\pi$ for which at specific values of the magnetic field $h$ over its saturation value $h_{sat}$ new magnetization or susceptibility discontinuities appear or disappear (Figs. \ref{fig:icosahedronconfigurations} and \ref{fig:icosahedronconfigurationsfivedifferent}). $J$ is the bilinear and $J'$ the biquadratic exchange interaction. Magnetization discontinuities are characterized by a number, and susceptibility discontinuities by a primed number. $N_M$ and $N_{\chi}$ give the total number of magnetization and susceptibility discontinuities right after discontinuities appear or disappear. The saturation magnetic field is $h_{sat}=(5+\sqrt{5})(cos\omega+2sin\omega)$ \cite{NPK15-1,comment}.}
\begin{tabular}{c|c|c|c|c|c|c|c|c|c}
 app. & disapp. & $\omega (\pi)$ & $\frac{h}{h_{sat}}$ & $N_M,N_{\chi}$ & app. & disapp. & $\omega (\pi)$ & $\frac{h}{h_{sat}}$ & $N_M,N_{\chi}$ \\
\hline
  1 & - & 0 & 0.40603 & 1,0 & 11 & 7,9 & 0.23178 & 0.16928 & 6,1 \\
\hline
  2,3 & 1 & 0.068897 & 0.29855 & 2,0 & 12 & 6,11 & 0.24366 & 0.15153 & 5,1 \\
\hline
  4 & - & 0.137    & 1 & 3,0 & 13 & 4,10 & 0.24535 & 0.38726 & 4,1 \\
\hline
  $1'$,$2'$ & - & 0.1695 & 0.183 & 3,2 & - & $4'$ & 0.2505 & 0.146 & 4,0 \\
\hline
  $3'$ & - & 0.1700 & 0.225 & 3,3 & 14 & 2,5 & 0.25328 & 0.046426 & 3,0 \\
\hline
  - & $2'$,$3'$ & 0.1701 & 0.218 & 3,1 & 15,16 & - & 0.26607 & 0.052924 & 5,0 \\
\hline
  5,6 & - & 0.18519  & 0.12489 & 5,1 & - & 14 & 0.28565 & 0 & 4,0 \\
\hline
  7,8 & 3 & 0.19061  & 0.21716 & 6,1 & 17,18 & 15 & 0.341601 & 0.029545 & 5,0 \\
\hline
  - & $1'$ & 0.2012 & 0.0894 & 6,0 & - & 16,18 & 0.341607 & 0.029652 & 3,0 \\
\hline
  9,10 & 8 & 0.20605 & 0.22171 & 7,0 & 19,20 & 17 & 0.48818 & 0.0030591 & 4,0 \\
\hline
  $4'$ & - & 0.2306 & 0.171  & 7,1 & - & 12,19,20 & $\frac{1}{2}$ & 0 & 1,0 \\
\end{tabular}
\label{table:magnsuscdisc}
\end{center}
\end{table}

\newpage

\end{document}